\documentclass[preprint,12pt]{elsarticle}

\usepackage{amssymb}

\usepackage{graphicx}
\usepackage{amsmath}
\usepackage{subfigure}

\def\jnl@style{\it}
\def\ref@jnl#1{{\jnl@style#1}}

\def\aj{\ref@jnl{AJ}}                   
\def\actaa{\ref@jnl{Acta Astron.}}      
\def\araa{\ref@jnl{ARA\&A}}             
\def\apj{\ref@jnl{ApJ}}                 
\def\apjl{\ref@jnl{ApJ}}                
\def\apjs{\ref@jnl{ApJS}}               
\def\ao{\ref@jnl{Appl.~Opt.}}           
\def\apss{\ref@jnl{Ap\&SS}}             
\def\aap{\ref@jnl{A\&A}}                
\def\aapr{\ref@jnl{A\&A~Rev.}}          
\def\aaps{\ref@jnl{A\&AS}}              
\def\azh{\ref@jnl{AZh}}                 
\def\baas{\ref@jnl{BAAS}}               
\def\bac{\ref@jnl{Bull. astr. Inst. Czechosl.}}
\def\caa{\ref@jnl{Chinese Astron. Astrophys.}}
\def\cjaa{\ref@jnl{Chinese J. Astron. Astrophys.}}
\def\GReGr{\ref@jnl{Gen. Rel. Grav.}}   
\def\icarus{\ref@jnl{Icarus}}           
\def\IJMPD{\ref@jnl{Int. J. Modern Phys. D}}    
\def\jcap{\ref@jnl{J. Cosmology Astropart. Phys.}}
\def\jrasc{\ref@jnl{JRASC}}             
\def\memras{\ref@jnl{MmRAS}}            
\def\mnras{\ref@jnl{MNRAS}}             
\def\na{\ref@jnl{New A}}                
\def\nar{\ref@jnl{New A Rev.}}          
\def\pla{\ref@jnl{Phys. Lett. A}}       
\def\plb{\ref@jnl{Phys. Lett. B}}       
\def\pra{\ref@jnl{Phys.~Rev.~A}}        
\def\prb{\ref@jnl{Phys.~Rev.~B}}        
\def\prc{\ref@jnl{Phys.~Rev.~C}}        
\def\prd{\ref@jnl{Phys.~Rev.~D}}        
\def\pre{\ref@jnl{Phys.~Rev.~E}}        
\def\prl{\ref@jnl{Phys.~Rev.~Lett.}}    
\def\pasa{\ref@jnl{PASA}}               
\def\pasp{\ref@jnl{PASP}}               
\def\pasj{\ref@jnl{PASJ}}               
\def\rvmp{\ref@jnl{Rev. Mod. Phys.}}    
\def\rmxaa{\ref@jnl{Rev. Mexicana Astron. Astrofis.}}%
\def\qjras{\ref@jnl{QJRAS}}             
\def\skytel{\ref@jnl{S\&T}}             
\def\solphys{\ref@jnl{Sol.~Phys.}}      
\def\sovast{\ref@jnl{Soviet~Ast.}}      
\def\ssr{\ref@jnl{Space~Sci.~Rev.}}     
\def\zap{\ref@jnl{ZAp}}                 
\def\nat{\ref@jnl{Nature}}              
\def\iaucirc{\ref@jnl{IAU~Circ.}}       
\def\aplett{\ref@jnl{Astrophys.~Lett.}} 
\def\apspr{\ref@jnl{Astrophys.~Space~Phys.~Res.}}
\def\bain{\ref@jnl{Bull.~Astron.~Inst.~Netherlands}} 
\def\fcp{\ref@jnl{Fund.~Cosmic~Phys.}}  
\def\gca{\ref@jnl{Geochim.~Cosmochim.~Acta}}   
\def\grl{\ref@jnl{Geophys.~Res.~Lett.}} 
\def\jcp{\ref@jnl{J.~Chem.~Phys.}}      
\def\jgr{\ref@jnl{J.~Geophys.~Res.}}    
\def\jqsrt{\ref@jnl{J.~Quant.~Spec.~Radiat.~Transf.}}
\def\memsai{\ref@jnl{Mem.~Soc.~Astron.~Italiana}}
\def\nphysa{\ref@jnl{Nucl.~Phys.~A}}   
\def\physrep{\ref@jnl{Phys.~Rep.}}   
\def\physscr{\ref@jnl{Phys.~Scr}}   
\def\planss{\ref@jnl{Planet.~Space~Sci.}}   
\def\procspie{\ref@jnl{Proc.~SPIE}}   

\journal{Physics Letters B}

\begin{document}

\begin{frontmatter}



\title{Constraining dark energy using observational growth rate data}


\author{K. Shi}
\address{Department of Astronomy, Nanjing University, Nanjing 210093, China}
\address{Key Laboratory of Modern Astronomy and Astrophysics (Nanjing University), Ministry of Education, Nanjing 210093, China}

\author{Y. F. Huang\corref{cor1}}
\ead{hyf@nju.edu.cn}
\address{Department of Astronomy, Nanjing University, Nanjing 210093, China}
\address{Key Laboratory of Modern Astronomy and Astrophysics (Nanjing University), Ministry of Education, Nanjing 210093, China}

\author{T. Lu}
\address{Purple Mountain Observatory, Chinese Academy of Sciences, Nanjing 210008, China}
\address{Joint Center for Particle, Nuclear Physics and Cosmology, Nanjing University -- Purple Mountain Observatory, Nanjing  210093, China}

\cortext[cor1]{Corresponding author}

\begin{abstract}
Observational growth rate data had been derived from observations of redshift distortions in 
galaxy redshift surveys. Here we use the growth rate data to place constraints on the dark energy 
model parameters. By performing a joint analysis with the Type Ia supernova, baryon acoustic 
oscillation and cosmic microwave background data, it is found that the growth rate data are 
useful for improving the constraints. The joint constraints show that the $\Lambda$CDM model is 
still in good agreement with current observations, although a time-variant dark energy still 
cannot be ruled out. It is argued that the growth rate data are helpful for understanding the 
dark energy. With more accurate data available in the future, we will have a powerful tool for 
constraining the cosmological and dark energy parameters. 
\end{abstract}


\end{frontmatter}


\section{Introduction}




The present accelerating expansion of the universe is a great challenge to our 
understanding of fundamental physics and cosmology. This fact was first revealed   
by Type Ia supernova (SNIa) surveys \cite{1998AJ....116.1009R,1999ApJ...517..565P}, and later confirmed by the precise measurement of the Cosmic Microwave Background (CMB) anisotropies \cite{2003ApJS..148..175S} as well as the baryon acoustic oscillations (BAO) in the Sloan Digital Sky Survey (SDSS) photometric galaxy sample \cite{2005ApJ...633..560E}. This cosmic acceleration leads us to believe that most energy in the universe exists in the form of a new ingredient called ``dark energy'', which has a negative pressure (see \cite{2003RvMP...75..559P, 2006IJMPD..15.1753C} for reviews).

Various theoretical models of dark energy have been proposed. 
The simplest one takes the form of a cosmological constant $\Lambda$, with a constant dark energy density and the equation of state $w_{DE} = p/\rho = -1$. This model, 
named the $\Lambda$CDM model, provides an excellent fit to a wide range of observations available so far. Despite its simplicity and success, the 
$\Lambda$CDM model has two major problems. One is the so-called ``fine tuning'' problem, 
i.e., the observed value of $\Lambda$ is extremely small as compared with the expections of particle physics  \cite{1989RvMP...61....1W}. The other is the coincidence problem. That is, the present densities 
of the dark energy ($\Omega_{\Lambda0}$) and matter ($\Omega_{m0}$) are of the same order of magnitude, 
for no obvious reasons. In order to solve these problems, alternative models have been proposed, 
including the dark energy scalar field models with a time varying dark energy density and equation of state. Examples include the quintessence model which has $w > -1$ \cite{1998PhRvL..80.1582C, 1999PhRvL..82..896Z}, 
and more exotic ``phantom'' models with $w < -1$ \cite{2002PhLB..545...23C}.

Although most studies show that the $\Lambda$CDM model is in good agreement with observational data, 
dynamical dark energy models cannot be excluded yet. In order to distinguish between different 
dark energy models, the most commonly used method is to constrain the dark energy equation of state $w$. Recent studies have already greatly improved the constraints on $w$. For example, 
the Supernova Legacy Survey three year sample (SNLS3), combining with a few other probes, 
indicated that $w$ = $-$ 1.061 $\pm$ 0.068 \cite{2011arXiv1104.1444S}. It should be noted 
that although these results are well consistent with the $\Lambda$CDM model, we still cannot 
determine whether the density of the dark energy is actually constant, or whether it varies 
with time as suggested by dynamical dark energy models.

SNIa is the first cosmological probe to present direct evidence for the existence of dark energy. 
Till now, it is still the most powerful tool to study the properties of dark energy. 
However, SNIa alone cannot give tight constraints on dark energy model parameters unless combined with other probes such as CMB and BAO and so on. As mentioned above, Sullivan et al. have integrated the CMB, 
BAO and Hubble constant data in their studies \cite{2011arXiv1104.1444S}. Indeed, the precision of modern cosmology lies in ``joint analysis'' - using different, independent observations rather than a single observational data set to obtain the final results. Therefore, every independent method is useful and important to help us understand cosmic acceleration.

Besides the SNIa, BAO and CMB data, many other observations have also been used to constrain the cosmological parameters. 
They include the Hubble parameter \cite{2010JCAP...02..008S}, gas mass fractions in galaxy clusters\cite{2008MNRAS.383..879A}, 
gamma-ray bursts  \cite{2007ApJ...660...16S, 2010ApJ...714.1347S}, large-scale structures \cite{2011MNRAS.410.1911C}, 
weak gravitational lensing \cite{2008A&A...479....9F}, strong gravitational lensing \cite{2010MNRAS.406.1055B}, 
and the lookback time  \cite{2004PhRvD..70l3501C}. While these data provide weaker constraints than the SNIa, BAO and CMB data do, they are independent and complementary. What surprises us is that 
they all generally support a currently accelerating expansion of the universe. This provides additional support to  
the $\Lambda$CDM model and leads us to believe that the observational data do not strongly mislead us despite their errors.

Large scale matter density perturbations in the universe are gravitationally unstable. They should grow  
according to linear theory. The growth of density perturbations can be used as another method to probe the 
dark energy as well as the modified gravity theory 
(see, e.g. \cite{2005PhRvD..72d3529L, 2006ApJ...648..797B, 2008Natur.451..541G}). 
The cosmic growth history is complementary to the cosmic expansion history. What is more, 
different dark energy models may predict similar expansion behavior, but the growth history could 
be different. So the growth rate of large scale structure is a powerful tool to distinguish between 
different dark energy models. There have already been a lot of work using the growth rate of large scale 
structures to constrain the dark energy and modified gravity theory 
(e.g. \cite{2008PhRvD..77b3504N, 2008PhRvD..77h3508D, 2010JCAP...11..004G, 2009PhRvD..79j3527X, 2011arXiv1103.6133H}). 
However, due to the sparsity of data and large error bars, previous growth rate data may not be robust and should 
be combined with other data sets. 

Recently, the WiggleZ Dark Energy Survey group has published a batch of observational growth rate 
data \cite{2010MNRAS.406..803B, 2011MNRAS.tmp..834B}. These data were obtained through a very rigorous analysis. 
They are consequently more precise and robust than previous data used in the literature. 
In this paper we will use these WiggleZ growth rate data along with earlier data gathered from the 
literature to set constraints on the cosmological parameters and the dark energy properties. 
We also perform joint analysises of these data with the SNIa, BAO and CMB data to get much tighter 
constraints on different dark energy models.

Our paper is organized as follows. In Section 2, we summarize the dark energy models to be tested 
in the study. In Section 3, we describe the growth rate data and the way to use them. 
In Section 4, the method of combining the SNIa, BAO and CMB data are described, which may hopefully 
provide much tighter constraints on the dark energy parameters. We present our final results in 
Section 5 and give a brief discussion in the last section.

\section{Dark energy models}
 
In this work we mainly consider three dark energy models. The first one is the most 
well-known $\Lambda$CDM model, for which the Friedmann equation is 
\begin{equation}
H^2(z) = H_0^2 [\Omega_{m} (1+z)^3 + \Omega_{\Lambda} + \Omega_k (1+z)^2],
\end{equation}
where $H(z) \equiv \dot{a}/a$ is the Hubble parameter, $\Omega_{m}$ is the current value of the normalised matter density, $\Omega_{\Lambda}$ represents the cosmological constant density and $\Omega_k=1-\Omega_m-\Omega_{\Lambda}$ is the present curvature density. There are two free parameters in this model -- $\Omega_m$ and $\Omega_{\Lambda}$.

In the $\Lambda$CDM model, the equation of state of the dark energy $w$ is fixed to be $-1$. To go 
a little further, we can consider $w$ as a free parameter to be fitted from observational data. 
In this wCDM model, for simplicity, the spatial curvature is usually set to be zero. The 
corresponding Friedmann equation is 
\begin{equation}
 H^2(z) = H_0^2 [\Omega_{m} (1+z)^3 + (1 - \Omega_{m}) (1+z)^{3(1+w)}],
\end{equation}
where the free parameters are $\Omega_m$ and $w$.

There is no prior reason to expect $w$ to be -1 or a constant. Actually, many function forms 
of $w$ evolving with redshift have been proposed so far (see, e.g. \cite{2007IJMPD..16.1581J}). 
Among the various parametrizations of the dark energy equation of state $w$, the one developed by Chevallier $\&$ Polarski \cite{2001IJMPD..10..213C} and Linder \cite{2003PhRvL..90i1301L} turns 
out to be an excellent approximation to a wide variety of dark energy models. Since this CPL 
(Chevalier-Polarski-Linder) model is the most commonly used function form for studying the  time dependence of $w$, we will examine it in this work with the new growth rate data. 
The equation of state in this model is 
\begin{equation}
w(z) = w_0 + w_a\frac{z}{1+z},
\end{equation}
where $w_0$ and $w_a$ are free parameters to be fitted from observational data. 
The curvature term is also set to be zero in this case, so that the Friedmann equation can be written as
\begin{equation}
H^2(z) = H_0^2 [\Omega_{m} (1+z)^3 + (1 - \Omega_{m}) (1+z)^{3(1+w_0+w_a)} \exp (\frac{-3 w_a z}{1+z})].
\end{equation}
There are three parameters in this model, $\Omega_m$, $w_0$ and $w_a$.

\section{Growth rate of matter perturbations}
In the case of linear cosmological perturbations, the equation governing the evolution of matter 
density fluctuations in an expanding universe is well-known:
\begin{equation}
 \ddot{\delta} + 2H\dot{\delta} = 4\pi G\rho_m \delta ,
\end{equation}
where $\delta=\delta \rho_m/\rho_m$ is the matter density perturbation and dots indicate time derivatives. 
If we define the linear growth rate of matter perturbations as $f \equiv d \ln \delta/d \ln a$, then Eq. (5) can be written as
\begin{equation}
f^2 + \frac{d f}{d \ln a} + (\frac{\dot{H}}{H^2} + 2) f = \frac{3}{2} \Omega_{m}(z),
\end{equation}
where $\Omega_{m}(z)=\Omega_{m}(1+z)^3/(H/H_0)^2$ is the dimensionless matter density as a function of redshift.

This growth rate factor can be well approximated as \cite{1980lssu.book.....P}
\begin{equation}
f(z) = \Omega_{m}^\gamma(z),
\end{equation}
where $\gamma$ is the growth index. Hence we have
\begin{equation}
\delta(z) = \delta(z_i) \exp{\int_z^{z_i} \Omega_{m}(z)^\gamma(z') \frac{dz'}{1+z'}}.
\end{equation}
In the $\Lambda$CDM model, the growth index is $\gamma \approx 0.55$, while in other dark energy 
and modified gravity models $\gamma$ should change correspondingly. It has been shown that for a 
wide range of dark energy models, $\gamma$ can be fitted to a high level of accuracy by \cite{2005PhRvD..72d3529L}
\begin{equation}
\gamma = 0.55 + 0.05[1 + w(z=1)].
\end{equation}

In our study, we use Eq. (7) and Eq. (9) to calculate the theoretical value of the growth rate $f(z)$. 
So in the wCDM model, the growth index becomes $\gamma=0.55+0.05(1+w)$, while for the CPL model, 
$\gamma=0.55+0.05(1+w_0+0.5w_a)$. There is another advantage in using the growth rate data, which is, 
we do not need to have any prior knowledge of the nuisance Hubble constant $H_0$. 

Now we present the observational growth rate data used in this paper (see Table 1). They were 
obtained from galaxy redshift surveys. Note that these values of $f$ were not measured directly. 
In fact, the redshift surveys measure the redshifts of galaxies and provide their distribution in the 
redshift space. However, the inferred redshift distribution is distorted from the real galaxy 
distribution because of the peculiar motions of galaxies. This is the so-called ``redshift distortion''. 

According to the linear theory, the observed power spectrum in the redshift space is related to the real 
power spectrum through \cite{1987MNRAS.227....1K}
\begin{equation}
P_{redshift}(k) = P_{real}(k) (1 + \beta \mu ^2)^2,
\end{equation}
where the redshift distortion parameter $\beta$ is connected with the growth rate $f$ and 
galaxy bias $b$ (i.e. the ratio between the galaxy density contrast and the total mass contrast) 
through $\beta \equiv f/b$, and $\mu$ is the cosine of the angle between the line of sight and 
the wavevector $k$. A commonly used method to extract $\beta$ from redshift surveys is to expand the redshift space correlation function (i.e. the inverse Fourier transform of the power spectrum) using a base of spherical harmonics with the aid of Eq. (10) . The galaxy bias $b$, however, is a nuisance parameter which can be obtained by scaling the matter power spectrum through linear theory \cite{2008Natur.451..541G} or just marginalized over \cite{2011MNRAS.tmp..834B}. Once we have measured $\beta$ and $b$, we can obtain the observed growth rate value $f$ through $f=\beta b$. 

There is one more issue which should be addressed here. Most of the data in Table 1 are determined under the assumption of the concordance $\Lambda$CDM model. The problem is that if the $\Lambda$CDM model is incorrect, it would induce a distortion on the resulting correlation maps that adds to the effect of the peculiar velocities we aim to measure. This is called the Alcock-Paczynski (AP) effect 
\cite{1979Natur.281..358A}. So when using these data, we should be very careful about this AP effect. 
We discuss this issue a little more below.

Guzzo et al. \cite{2008Natur.451..541G} have used a set of simulations to test the ``model-dependent'' 
feature of $\beta$. To their surprise, they found that even if they had mistaken 
$\Omega_{m}$ by a large factor, they still could get a fairly correct value of $\beta$. Thus they concluded that their measured value of $\beta$ is robust against this AP distortion. In our work, 
since the models to be tested are not that far from the $\Lambda$CDM model, and furthermore, the redshifts of 
these growth rate data points are smaller compared to those of Guzzo et al. \cite{2008Natur.451..541G}, it follows that when converting the redshifts into the comoving distances to measure $f$, this 
model-dependent bias should be relatively small. The AP effect thus should be small and can be 
neglected in our study, as compared to the large errorbars of the observational data points of 
$f$. However, we caution that it is dangerous to use these growth rate data to constrain other 
dark energy models which deviate far from the $\Lambda$CDM model (e.g. the f(R) theory, the DGP 
model). 

In order to constrain the cosmological parameters from the growth rate data, 
we define the $\chi^2$ statistic as
\begin{equation}
\chi^2_f = \sum_{i=1}^{11} \frac{(f_{obs}(z_i)-f_{th}(z_i))^2}{\sigma_i ^2},
\end{equation}
where the theoretical value $f_{th}$ is calculated through Eqs. (7) and (9), and $\sigma_i$ is the 
corresponding observational uncertainty.

To get a sense of the dependence of the growth rate $f$ on different cosmological parameters, 
we plot the evolution behavior of $f$ as a function of redshift $z$ for different values of 
$\Omega_m$ and $w$ in Fig.~1. We can see that the influence of $w$ on the growth rate $f(z)$ 
is small at relatively low redshifts and becomes much larger at high redshifts. On the other 
hand, for the three values of $\Omega_m$ considered in Fig.~1, the curve of $f(z)$ deviate 
from each other. This is expected, as one can see from Eq.~(7) that the contribution of $\Omega_m$ comes from $\Omega_m(z)$ while $w$ mainly affects the index $\gamma$. The 11 
observational data points are also shown in the figure. It can be seen that the error bars 
are quite large, which is still a serious problem for the current growth rate data. However, 
it can still be seen that the middle panel coincides with the data very well, suggesting 
that $\Omega_m$ lies around 0.3, which is in agreement with other observational constraints. 
This preliminary result will be studied quantitatively in the next section.

\section{SNIa, BAO and CMB data}
As we have mentioned earlier, due to the large error bars of growth rate data, we need to 
integrate other data to obtain better constraints on the cosmological parameters. In this 
paper we use the SNIa, BAO and CMB data to place joint constraints on the dark energy models. 

\subsection{SNIa data}
Currently, SN Ia is the most powerful tool to study the dark energy due to its reliability 
as standard candles. For the SNIa data, we use the latest UNION2 compilation 
\cite{2010ApJ...716..712A} which, in total, contains 557 type Ia supernovae with redshifts 
ranging from 0.511 to 1.12. Since there are too many data points, we will not list 
them here. For details of the data, readers can refer to Amanullah et al. \cite{2010ApJ...716..712A}. 
Cosmological constraints from the SNIa data are obtained through 
the distance modulus $\mu(z)$. The theoretical distance modulus is
\begin{equation}
\mu_{th}(z_i) = 5 \log_{10} D_L(z_i) + \mu_0,
\end{equation}
where $\mu_0 = 42.38-5\log_{10}h$ with $h$ being the Hubble constant $H_0$ in units of 
100 km/s/Mpc, and the Hubble-free luminosity distance $D_L$ is defined as
\begin{equation}
D_L(z) = \frac{1+z}{\sqrt{\vert\Omega_k\vert}}\textrm{sinn}\left[\sqrt{\vert\Omega_k\vert}\int_0^z \frac{dz'}{E(z')}\right],
\end{equation}
where $E(z) = H(z)/H_0$ and $\Omega_k$ is the present curvature density. Here the symbol 
sinn(x) stands for sinh(x) ($\Omega_k>0$), sin(x) ($\Omega_k<0$) or just x ($\Omega_k=0$).

To compute the $\chi^2$ for the SNIa data, we follow \cite{2005PhRvD..72l3519N} to analytically 
marginalize over the nuisance parameter $H_0$, 
\begin{equation}
\chi_{SN}^2 = A - 2\mu_0 B + \mu_0^2C,
\end{equation}
where 
\begin{equation}
\begin{aligned}
&A = \sum_{i=1}^{557} \frac{[\mu_{obs}(z_i)-\mu_{th}(z_i;\mu_0=0)]^2}{\sigma_i^2},\\
&B = \sum_{i=1}^{557} \frac{\mu_{obs}(z_i)-\mu_{th}(z_i;\mu_0=0)}{\sigma_i^2},\\
&C = \sum_{i=1}^{557} \frac{1}{\sigma_i^2}.
\end{aligned}
\end{equation}
Here $\sigma$ is the uncertainty in the distance moduli. Eq.~(14) has a minimum at $\mu_0 = B/C$ where
\begin{equation}
\tilde{\chi}_{SN}^2 = A - \frac{B^2}{C}.
\end{equation}
This equation is independent of $\mu_0$, so instead of $\chi_{SN}^2$ we adopt $\tilde{\chi}_{SN}^2$ to compute the likelihood.

\subsection{BAO data}

The competition between the gravitational force and the relativistic primordial plasma pressure gives 
rise to acoustic oscillations which leaves its signature in every epoch of the universe. 
Eisenstein et al.  \cite{2005ApJ...633..560E} first found a peak of this baryon acoustic 
oscillations in the large-scale correlation function at 100 $h^{-1}$ Mpc separation 
from a spectroscopic sample of 46,748 luminous red galaxies of the Sloan Digital Sky 
Survey (SDSS).  Percival et al. \cite{2010MNRAS.401.2148P} performed a power-spectrum 
analysis of the SDSS DR7 dataset, considering both the main and LRG samples, and measured 
the BAO signal at both $z=0.2$ and $z=0.35$. For low redshifts, the 6dF Galaxy Survey (6dFGS) 
group have reported a BAO detection at $z=0.1$ \cite{2011MNRAS.416.3017B}. Most recently, 
Blake et al. \cite{2011MNRAS.tmp.1598B} presented measurements of the BAO peak at 
redshifts $z=0.44,0.6$ and 0.73 in the galaxy correlation function of the final dataset 
of the WiggleZ Dark Energy Survey. They combined their WiggleZ BAO measurements with the 
SDSS DR7 and 6dFGS datasets to present tight constraints on the dark energy. In this work, 
We follow their methods to constrain different dark energy models by using their combined BAO datasets. 

The data can be found in the above papers, but for completeness we summarize the BAO measurements 
from the three surveys and the way we use them.

The $\chi^2$ for the WiggleZ BAO data is given by \cite{2011MNRAS.tmp.1598B},
\begin{equation}
\chi^2_{\scriptscriptstyle WiggleZ} = (\bar{A}_{obs}-\bar{A}_{th}) C_{\scriptscriptstyle WiggleZ}^{-1} (\bar{A}_{obs}-\bar{A}_{th})^T,
\end{equation}
where the data vector is $\bar{A}_{obs} = (0.474,0.442,0.424)$ for the effective redshift $z=0.44,0.6$ and 0.73. The corresponding theoretical value $\bar{A}_{th}$ denotes the acoustic parameter $A(z)$ introduced by \cite{2005ApJ...633..560E}:
\begin{equation}
A(z) = \frac{D_V(z)\sqrt{\Omega_{m}H_0^2}}{cz},
\end{equation}
where the distance scale $D_V$ is defined as
\begin{equation}
D_V(z)=\frac{1}{H_0}\left[(1+z)^2D_A(z)^2\frac{cz}{E(z)}\right]^{1/3}.
\end{equation}
Here $D_A(z)$ is the Hubble-free angular diameter distance which relates to the Hubble-free luminosity distance through $D_A(z)=D_L(z)/(1+z)^2$, and $E(z)=H(z)/H_0$.
The inverse covariance matrix $C_{\scriptscriptstyle WiggleZ}^{-1}$ is given by
\begin{equation}
C_{\scriptscriptstyle WiggleZ}^{-1} = \left(
\begin{array}{ccc}
1040.3 & -807.5 & 336.8\\
-807.5 & 3720.3 & -1551.9\\
336.8 & -1551.9 & 2914.9
\end{array}\right).
\end{equation}

Similarly, for the SDSS DR7 BAO distance measurements, the $\chi^2$ can be expressed as \cite{2010MNRAS.401.2148P}
\begin{equation}
\chi^2_{\scriptscriptstyle SDSS} = (\bar{d}_{obs}-\bar{d}_{th})C_{\scriptscriptstyle SDSS}^{-1}(\bar{d}_{obs}-\bar{d}_{th})^T,
\end{equation}
where $\bar{d}_{obs} = (0.1905,0.1097)$ is a vector containing the datapoints at $z=0.2$ and $0.35$. $\bar{d}_{th}$ denotes the theoretical distance ratio 
\begin{equation}
d_z = \frac{r_s(z_d)}{D_V(z)}.
\end{equation}
Here, $r_s(z)$ is the comoving sound horizon,
\begin{equation}
 r_s(z) = c \int_z^\infty \frac{c_s(z')}{H(z')}dz',
 \end{equation}
where the sound speed $c_s(z) = 1/\sqrt{3(1+\bar{R_b}/(1+z)}$, with $\bar{R_b} = 31500 \Omega_{b}h^2(T_{CMB}/2.7\rm{K})^{-4}$ and $T_{CMB}$ = 2.726K.

The redshift $z_d$ at the baryon drag epoch is fitted with the formula of    \cite{1998ApJ...496..605E},
\begin{equation}
z_d = \frac{1291(\Omega_{m}h^2)^{0.251}}{1+0.659(\Omega_{m}h^2)^{0.828}}[1+b_1(\Omega_b h^2)^{b_2}],
\end{equation}
where
\begin{equation}
\begin{aligned}
&b_1 = 0.313(\Omega_{m}h^2)^{-0.419}[1+0.607(\Omega_{m}h^2)^{0.674}], \\
&b_2 = 0.238(\Omega_{m}h^2)^{0.223}.
 \end{aligned}
\end{equation}

$C_{\scriptscriptstyle SDSS}^{-1}$ in Eq. (21) is the inverse covariance matrix for the SDSS data set given by
\begin{equation}
C_{\scriptscriptstyle SDSS}^{-1} = \left(
\begin{array}{cc}
30124 & -17227\\
-17227 & 86977
\end{array}\right).
\end{equation}

For the 6dFGS BAO data \citep{2011MNRAS.416.3017B}, there is only one data point at $z=0.106$, so the $\chi^2$ is easy to compute:
\begin{equation}
\chi^2_{\scriptscriptstyle 6dFGS} = \left(\frac{d_z-0.336}{0.015}\right)^2.
\end{equation}

The total $\chi^2$ for all the BAO data sets can therefore be written as
\begin{equation}
\chi^2_{BAO} = \chi^2_{\scriptscriptstyle WiggleZ} + \chi^2_{\scriptscriptstyle SDSS} + \chi^2_{\scriptscriptstyle 6dFGS}.
\end{equation}
For clarity, we have given all the BAO data sets used in our study in Table 2. 
The errors are contained in the inverse covariance matrixes presented in Eqs.~(20) and (26).

\subsection{CMB data}

Since the SNIa and BAO data contain information about the universe at relatively low redshifts, 
we will include the CMB information by implementing the WMAP 7-year data \cite{2011ApJS..192...18K} 
to probe the entire expansion history up to the last scattering surface. 
The $\chi^2$ for the CMB data is constructed as
 \begin{equation}
 \chi^2_{CMB} = X^TC^{-1}X,
 \end{equation}
 where
 \begin{equation}
 X =\left(
 \begin{array}{c}
 l_A - 302.09 \\ 
 R - 1.725 \\
 z_* - 1091.3
 \end{array}\right).
 \end{equation}
 Here $l_A$ is the ``acoustic scale'' defined as
\begin{equation}
l_A = \frac{\pi d_L(z_*)}{(1+z)r_s(z_*)},
\end{equation}
where the redshift of decoupling $z_*$ is given by \cite{1996ApJ...471..542H},
\begin{equation}
z_* = 1048[1+0.00124(\Omega_bh^2)^{-0.738}] [1+g_1(\Omega_{m}h^2)^{g_2}],
\end{equation}
\begin{equation}
g_1 = \frac{0.0783(\Omega_bh^2)^{-0.238}}{1+39.5(\Omega_bh^2)^{0.763}}, 
 g_2 = \frac{0.560}{1+21.1(\Omega_bh^2)^{1.81}},
\end{equation}
The ``shift parameter'' ($R$) in Eq.~(30) is defined as \cite{1997MNRAS.291L..33B}
\begin{equation}
R = \frac{\sqrt{\Omega_{m}}}{c} \int_0^{z_*} \frac{dz}{E(z)}, 
\end{equation}
and $C^{-1}$ is the inverse covariance matrix
\begin{equation}
C^{-1} = \left(
\begin{array}{ccc}
2.305 & 29.698 & -1.333\\
29.698 & 6825.270 & -113.180\\
-1.333 & -113.180 & 3.414
\end{array}\right).
\end{equation}

For completeness, we have also given the CMB data used in our study in Table 2.
The errors are contained in the inverse covariance matrixes presented in Eq.~(35).

\section{Results}
To show the function of the growth rate data, we first use these data alone to constrain the 
three dark energy models described in Section 2. We then combine the growth rate data with the SNIa, 
BAO and CMB data to get the joint results. We use the Markov Chain Monte Carlo (MCMC) method to 
search in the parameter spaces and determine the best-fit parameters. Our MCMC code is based on 
the publicly available CosmoMC package \cite{2002PhRvD..66j3511L}.

Figs. 2$-$4 show the constraints on the dark energy models from the growth rate data alone.
It can be seen that the current growth rate data generally support the $\Lambda$CDM model 
within the 1$\sigma$ confidence level. However, the contours are quite broad and are not as stringent as those from the SNIa or BAO + CMB data. Anyway, the results are comparable to those 
of the Hubble parameter data \cite {2011arXiv1106.4294C}, 
the gamma-ray burst data \cite{2010ApJ...714.1347S}, 
and the strong gravitational lensing data \cite{2011arXiv1105.6226C}.  
The best fit parameters are listed in Table 3. We notice that the growth rate data alone tend 
to favor a larger $\Omega_m$ in the $\Lambda$CDM and wCDM case. The constraints on 
$\Omega_{\Lambda}$ and $w$ are not that stringent, mainly due to the sparsity and large error bars of current growth rate data.  As for the CPL model, because of the large number of parameters, 
the constraints are much weaker. Despite the large error bars of current growth rate data, it can be seen that they still give reasonable results for 
the three dark energy models, and the concordance $\Lambda$CDM model is still favored by these 
growth rate data. 

To get better constraints on the dark energy parameters and to see the influence on 
the results when adding the growth rate data to other data sets, we also perform a 
joint analysis combining the SNIa, BAO and CMB data. For this purpose, the total 
$\chi^2$ is defined as
\begin{equation}
\chi_{total}^2 = \chi_{f}^2 + \chi_{SN}^2 + \chi_{BAO}^2 +\chi_{CMB}^2.
\end{equation}

Figs. 5$-$7 show the constraints on the cosmological parameters with and without the growth rate data. 
It can be seen that adding the growth rate data to the combination of the SNIa, BAO and CMB data do affect 
the results. In all the three figures, including the growth rate data improves the joint constraints. 
This is most obvious in Fig.~5 and Fig.~7. In Fig.~6, the results are also improved slightly by adding 
the growth rate data. Meanwhile, the contours shift somewhat towards the high-w end, suggesting that the growth rate data prefer a larger $w$, consistent with the earlier results we just obtained. 
Of course, currently these growth rate data are not able to compete with the three most 
powerful datasets -- SNIa, CMB and BAO, but they can help strengthen the constraints on the dark energy, which is a good sign. Table 3 also presents the best-fit parameters of the three dark energy models 
using different combinations of the data sets. The results are consistent with those of Blake et al. 
\cite{2011MNRAS.tmp.1598B}, who have also tested the three dark energy models using the new SNIa, BAO and CMB data. 

Generally speaking, the concordance $\Lambda$CDM model is well consistent with currently available data. 
However, alternative models like the wCDM and CPL models cannot be excluded yet. Therefore, more 
observational data are required for a clear discrimination between these models. 
Anyway, our study shows that we can use the growth rate data as an important supplement to the SNIa, BAO and CMB data 
to probe the dark energy, as they can serve to strengthen the constraints. Currently these data are heavily 
restricted by the large errors. It is expected that future observations will provide more of these 
independent growth rate data along with higher accuracy, to help us place better constraints on the 
dark energy.

\section{Conclusion and discussion}
In this work, we have used the observational growth rate data to set constraints on 
three dark energy models. These growth rate data are mainly derived from redshift surveys 
using redshift distortions, including the latest WiggleZ dark energy survey data 
which are much more precise than previous data.

Because most of the data are obtained under the assumption of general relativity and 
the $\Lambda$CDM model, we do not use them to constrain the modified gravity models 
as usually done in previous studies. To be conservative, we only use them to constrain the mildly 
time-evolving dark energy models which deviate not quite far from the $\Lambda$CDM model. 
We expect that future redshift surveys would examine the AP effect more accurately, and thus 
could improve the usage of the growth rate data for constraining the cosmological parameters.
 
There is an important advantage in using these data to constrain the cosmological parameters. 
Unlike many other cosmological probes (e.g. the BAO, CMB and Observational Hubble parameter), 
the growth rate data are completely independent of the Hubble constant $H_0$, thus they may 
help break the degeneracy between $H_0$ and other parameters. Due to the large error bars and 
the sparsity of the available data, the growth rate data alone currently cannot give 
quite stringent constraints on the model parameters. But they still support the concordance 
$\Lambda$CDM model at the 1$\sigma$ confidence level, as shown in Figs. 2$-$4. 
Meanwhile, they seem to prefer larger values of $\Omega_m$ and $w$. Combined with other data sets, 
the growth rate data can improve the constraints, as can be seen in Figs.~5$-$7, where the contours 
become smaller after adding the growth rate data. Our joint analysises show that the 
$\Lambda$CDM model is still in good agreement with current data, but a time-evolving dark energy 
can not be excluded yet. As an independent data set, the growth rate data can help tighten the 
constraints of the SNIa, BAO and CMB data. It is a good sign for using the growth rate data 
to constrain the cosmological parameters.  

We anticipate that future redshift surveys will provide more reliable and accurate growth rate data. 
In conjunction with other observational data, the growth rate data should be quite helpful for 
us to understand the cosmic acceleration.

\section*{Acknowledgments}
We thank the anonymous referee for helpful comments and suggestions. KS is very grateful to Luigi Guzzo 
for valuable discussions and suggestions. We also would like to thank Shi Qi and Bo Yu for useful 
discussions. This work was supported by the National Basic Research Program of 
China (973 program, Grant No. 2009CB824800), the National Natural Science Foundation of China 
(Grant Nos. 11033002 and 10973039).

\begin{table*}
\begin{center}
\caption{The observational growth rate data.}
\begin{tabular}{cccc}
\hline \hline
z & f(z) & $\sigma$ & Reference \\
\hline
0.15 & 0.49 & 0.14 & \cite{2008Natur.451..541G}\\
0.22 & 0.60 & 0.10 & \cite{2011MNRAS.tmp..834B}\\
0.34 & 0.64 & 0.09 & \cite{2009MNRAS.393.1183C}\\
0.35 & 0.70 & 0.08 & \cite{2006PhRvD..74l3507T}\\
0.41 & 0.70 & 0.07 & \cite{2011MNRAS.tmp..834B}\\
0.42 & 0.73 & 0.09 & \cite{2010MNRAS.406..803B}\\
0.55 & 0.75 & 0.18 & \cite{2007MNRAS.381..573R}\\
0.59 & 0.75 & 0.09 & \cite{2010MNRAS.406..803B}\\
0.60 & 0.73 & 0.07 & \cite{2011MNRAS.tmp..834B}\\
0.77 & 0.91 & 0.36 & \cite{2008Natur.451..541G}\\
0.78 & 0.70 & 0.08 & \cite{2011MNRAS.tmp..834B}\\ 
\hline
\hline 
\end{tabular}
\end{center}
\end{table*}

\begin{table*}
\begin{center}
\caption{The BAO and CMB distance data sets}
\begin{tabular}{ccccc}
\hline \hline
BAO & z & $d_z$ & A(z) & Reference\\
\hline
6dFGS & 0.106 & 0.336 & & \cite{2011MNRAS.416.3017B}\\
SDSS & 0.2 & 0.1905 & & \cite{2010MNRAS.401.2148P}\\
SDSS & 0.35 & 0.1097 & & \cite{2010MNRAS.401.2148P}\\

WiggleZ & 0.44 & & 0.474 & \cite{2011MNRAS.tmp.1598B}\\
WiggleZ & 0.6 & & 0.442 & \cite{2011MNRAS.tmp.1598B}\\
WiggleZ & 0.73 & & 0.424 & \cite{2011MNRAS.tmp.1598B}\\
\hline
CMB & $z_*$ & $l_A$ & R & Reference\\
\hline
WMAP & 1091.3 & 302.09 & 1.725 & \cite{2011ApJS..192...18K}\\
\hline 
\hline
\end{tabular}
\end{center}
\end{table*}

\begin{table*}
\begin{center}
\caption{ Best-fit parameters (1$\sigma$) of the three dark energy models using different  combinations of data sets.}
\begin{tabular}{cccc}
\hline \hline
Model & growth rate & SNIa + BAO + CMB & SNIa + BAO + CMB + growth rate\\
\hline
$\Lambda$CDM & $\Omega_{m}$ = $0.330 \pm 0.033$ & $\Omega_{m}$ = $0.292 \pm 0.007$ & $\Omega_{m}$ = $0.287 \pm 0.006$ \\
& $\Omega_{\Lambda} = 0.421^{+0.130}_{-0.421}$ & $\Omega_{\Lambda} = 0.713 \pm 0.007$ & $\Omega_{\Lambda} = 0.717 \pm 0.007$\\
\hline
wCDM & $\Omega_{m}$ = $0.342 \pm 0.110$ & $\Omega_{m}$ = $0.292 \pm 0.006$ & $\Omega_{m}$ = $0.290 \pm 0.007$ \\
 & $w$ = $-0.715^{+0.407}_{-0.400}$ & $w$ = $-0.982 \pm 0.037$ & $w$ = $-0.953 \pm 0.031$ \\
\hline
CPL & $\Omega_{m}$ = $0.280^{+0.187}_{-0.176}$ & $\Omega_m$ = $0.292 \pm 0.015$ & $\Omega_m$ = $0.289 \pm 0.015$ \\
& $w_0$ = $-1.911^{+2.377}_{-2.436}$ & $w_0$ = $-0.999 \pm 0.172$ & $w_0 = -0.997 \pm 0.157$ \\
 & $w_a$ = $4.054^{+8.119}_{-7.963}$ & $w_a$ = $0.385^{+0.271}_{-0.293}$ & $w_a$ = $0.484^{+0.239}_{-0.290}$ \\
 \hline \hline

\end{tabular}
\end{center}
\end{table*}

\begin{figure*}
\centering
\subfigure{\includegraphics[width=4.0in,height=3.0in]{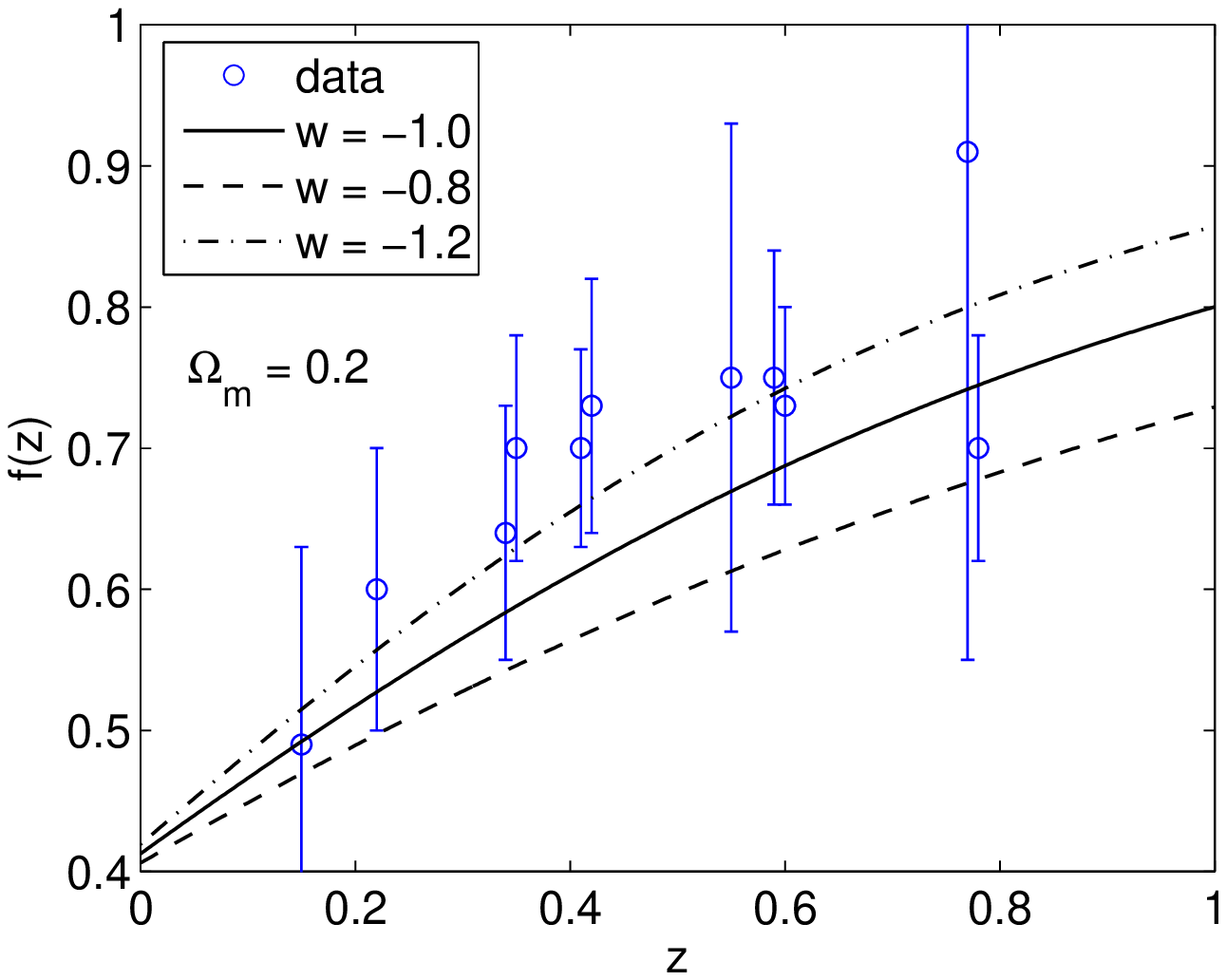}}
\subfigure{\includegraphics[width=4.0in,height=3.0in]{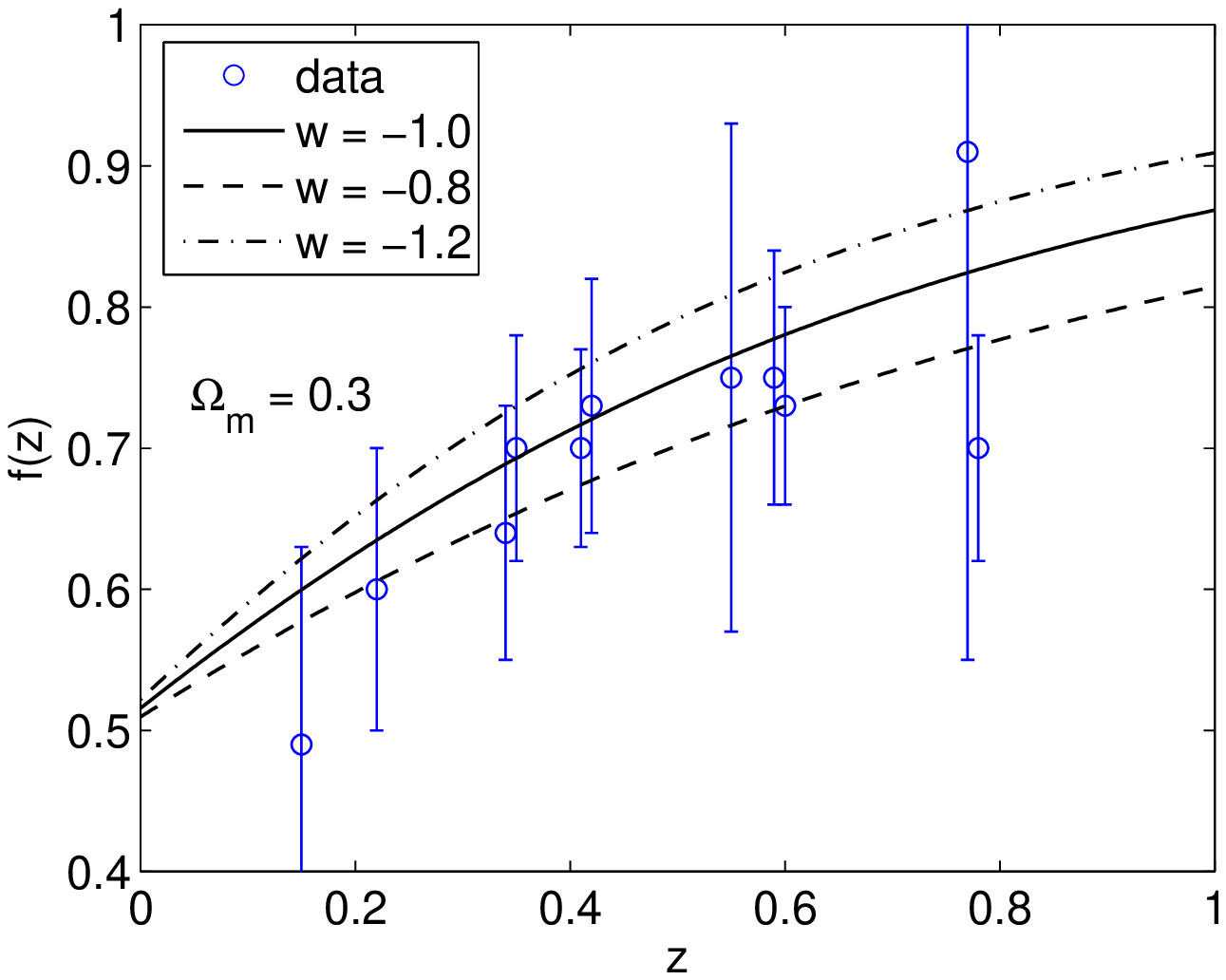}}
\subfigure{\includegraphics[width=4.0in,height=3.0in]{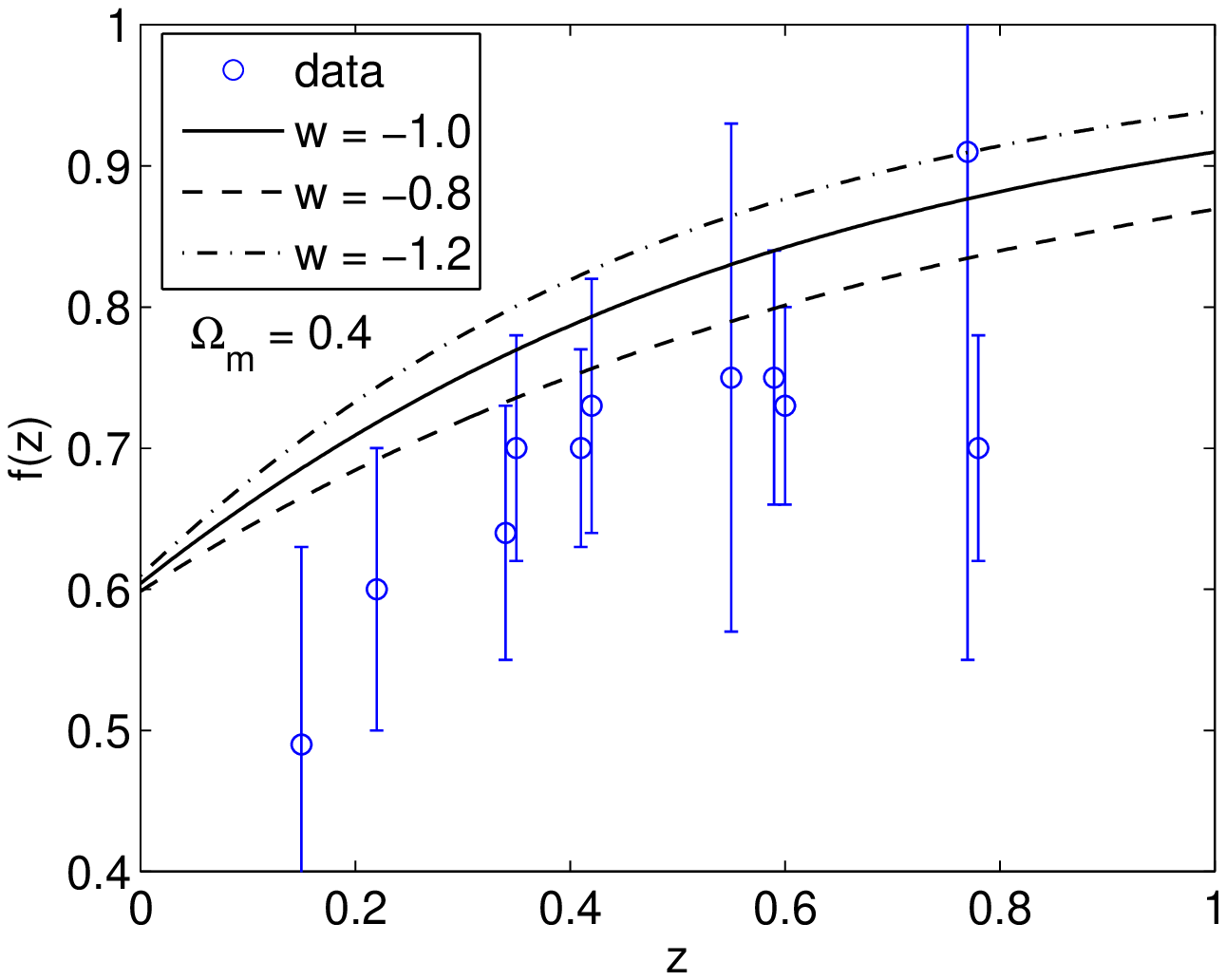}}
\caption{The evolution behavior of the growth rate $f(z)$ for different values of $\Omega_m$ and $w$. 
The observational data are also plotted with their errorbars.}
\end{figure*}

\begin{figure*}
\includegraphics[width=7in]{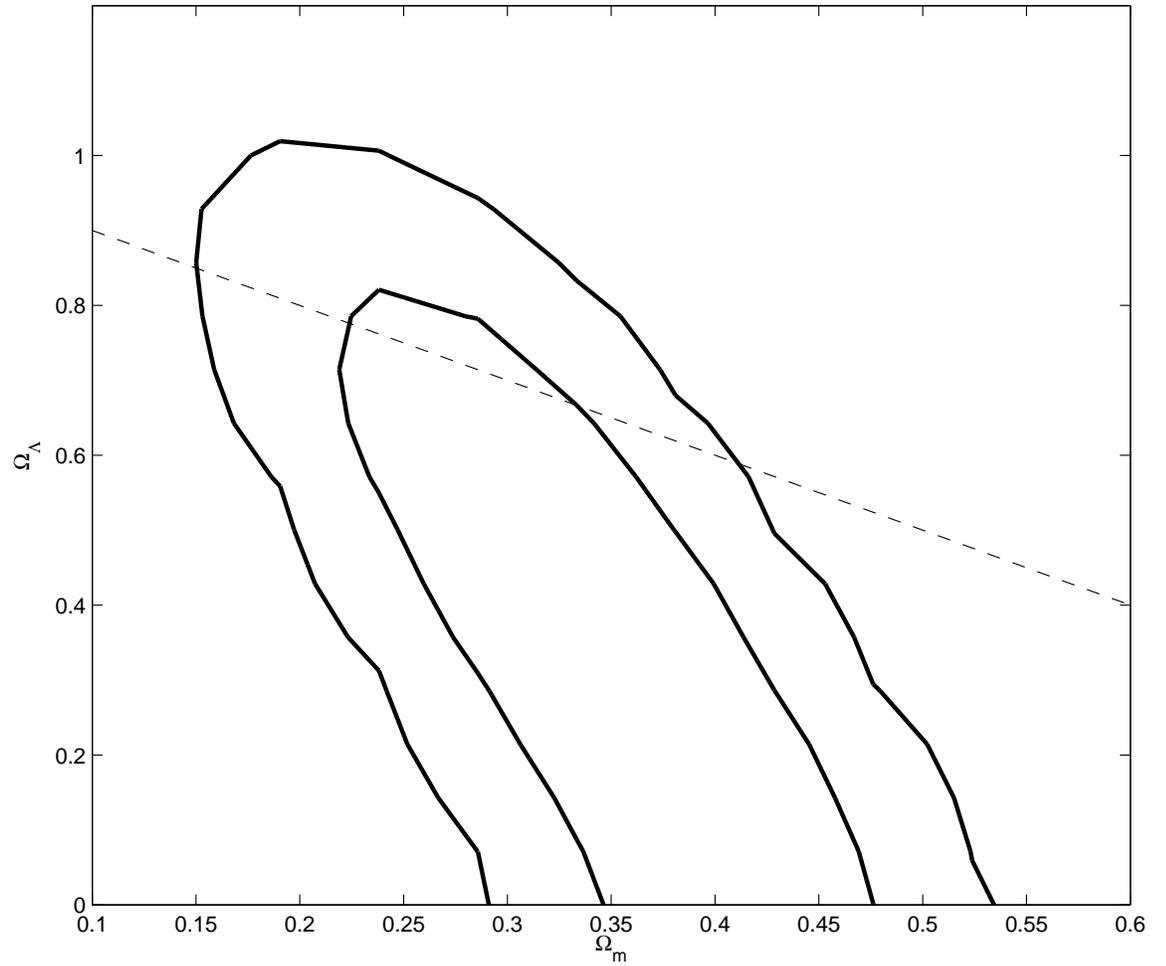}
\caption{1$\sigma$ and 2$\sigma$ $\Omega_{m}$-$\Omega_{\Lambda}$ contours for the $\Lambda$CDM model 
by using the growth rate data alone. The dashed line corresponds to a spacially flat universe.}
\end{figure*}

\begin{figure*}
\includegraphics[width=7in]{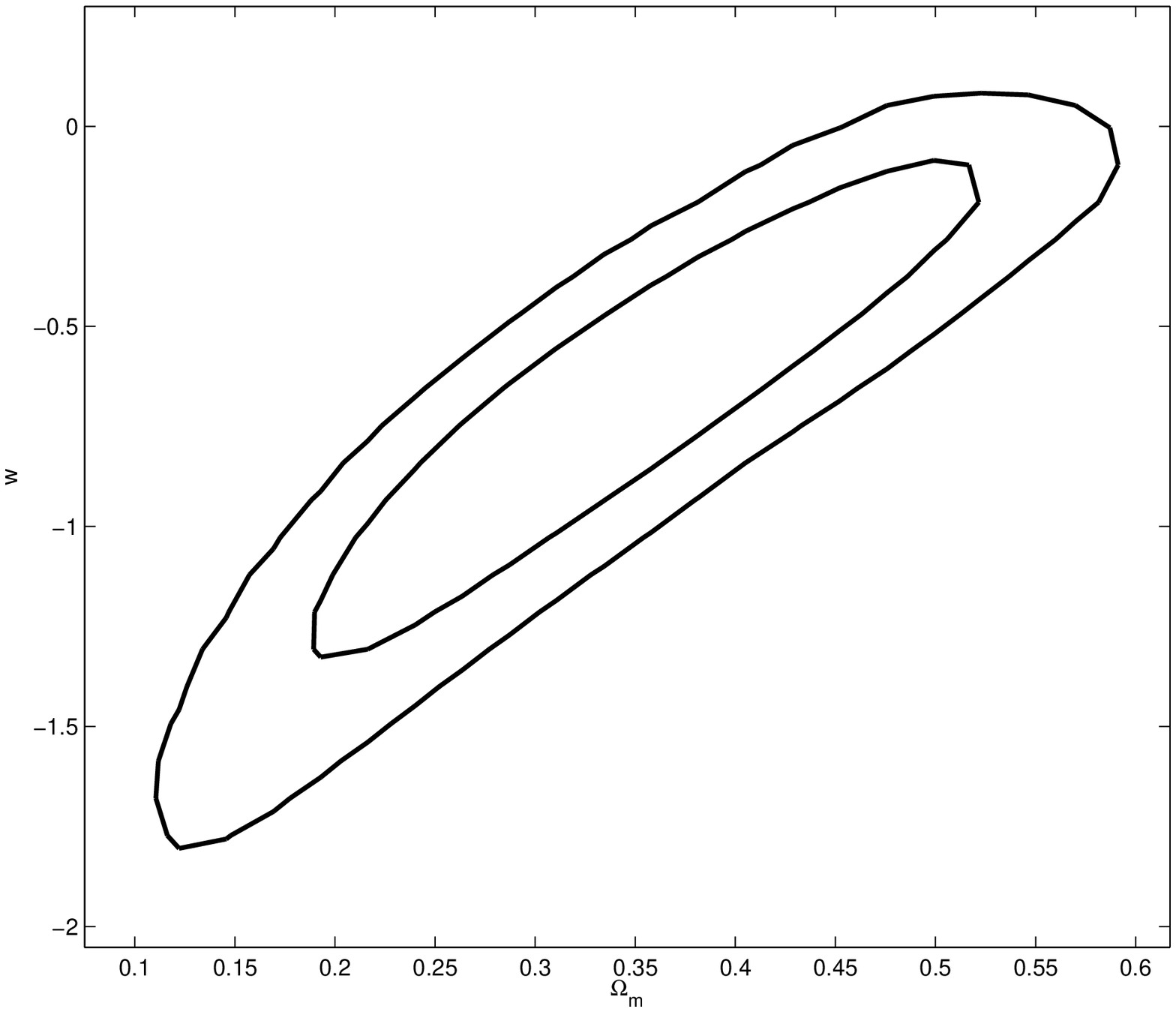}
\caption{1$\sigma$ and 2$\sigma$ $\Omega_{m}$-$w$ contours for the wCDM model obtained by using
the growth rate data alone.}
\end{figure*}

\begin{figure*}
\includegraphics[width=7in]{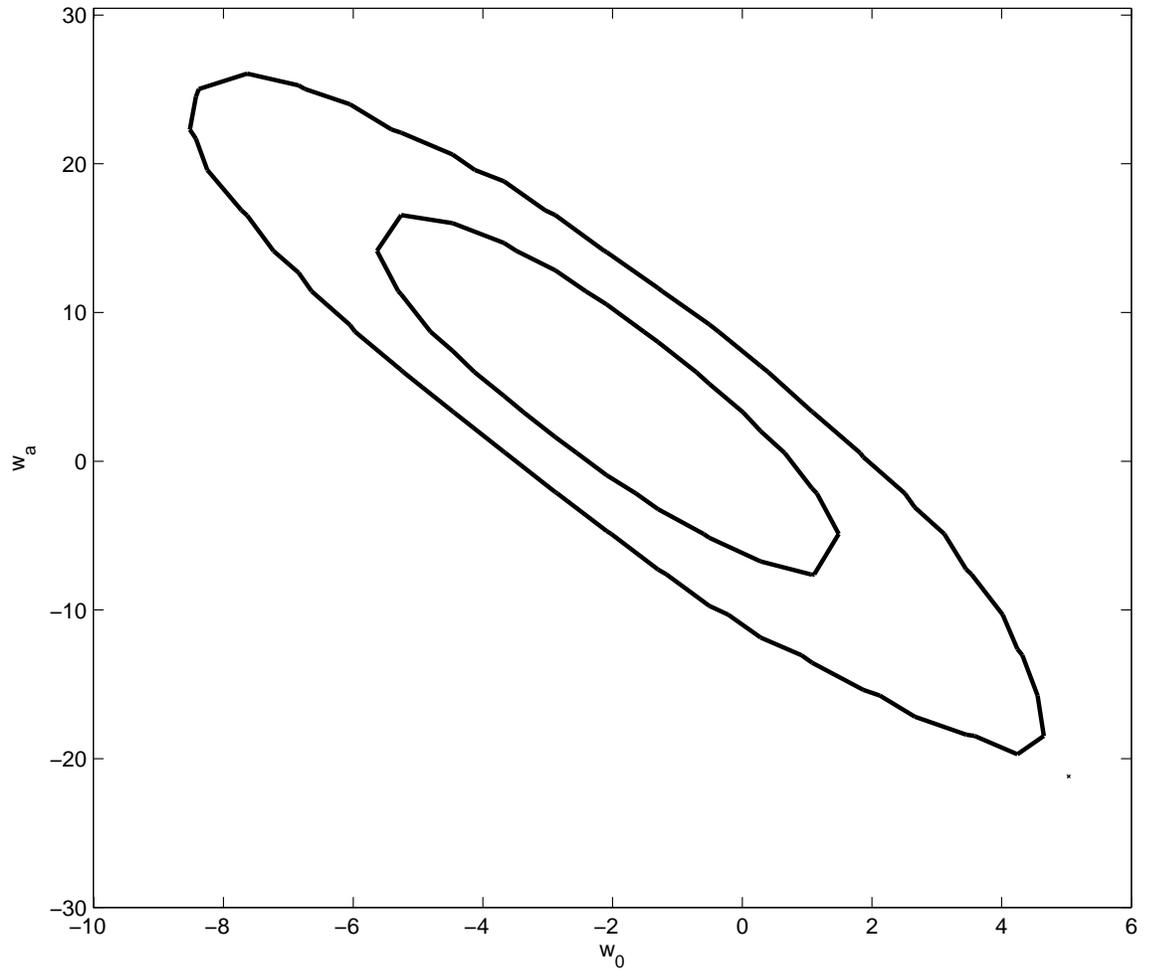}
\caption{1$\sigma$ and 2$\sigma$ $w_0$-$w_a$ contours for the CPL model obtained by 
using the growth rate data alone.}
\end{figure*}

\begin{figure*}
\includegraphics[width=7in]{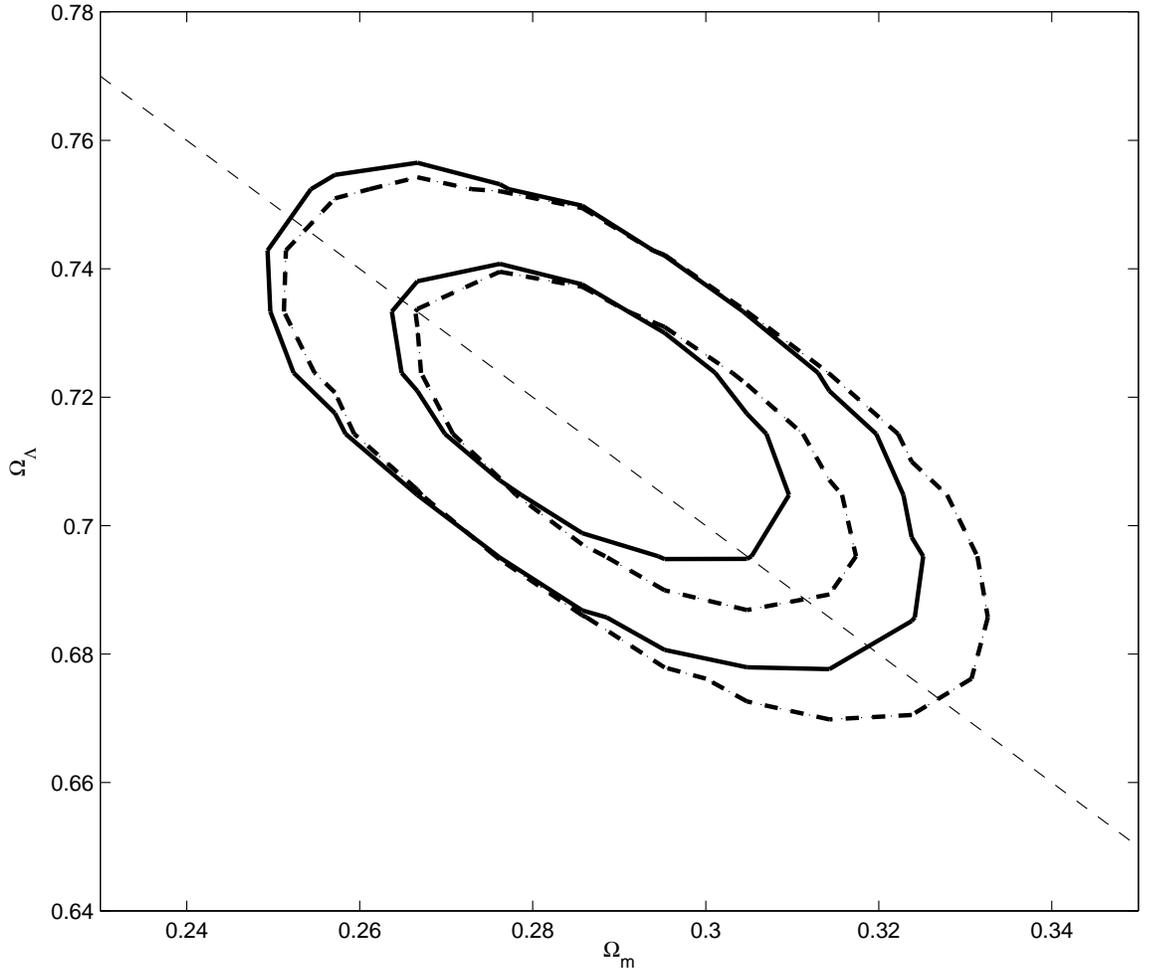}
\caption{1$\sigma$ and 2$\sigma$ $\Omega_{m}$-$\Omega_{\Lambda}$ contours for 
the $\Lambda$CDM model. The solid lines correspond to the result of a combined analysis of the 
growth rate, SNIa, BAO and CMB data, while the dash-dotted lines are obtained using only the 
SNIa, BAO and CMB data. The dashed straight line corresponds to a spacially flat universe.}
\end{figure*} 

\begin{figure*}
\includegraphics[width=7in]{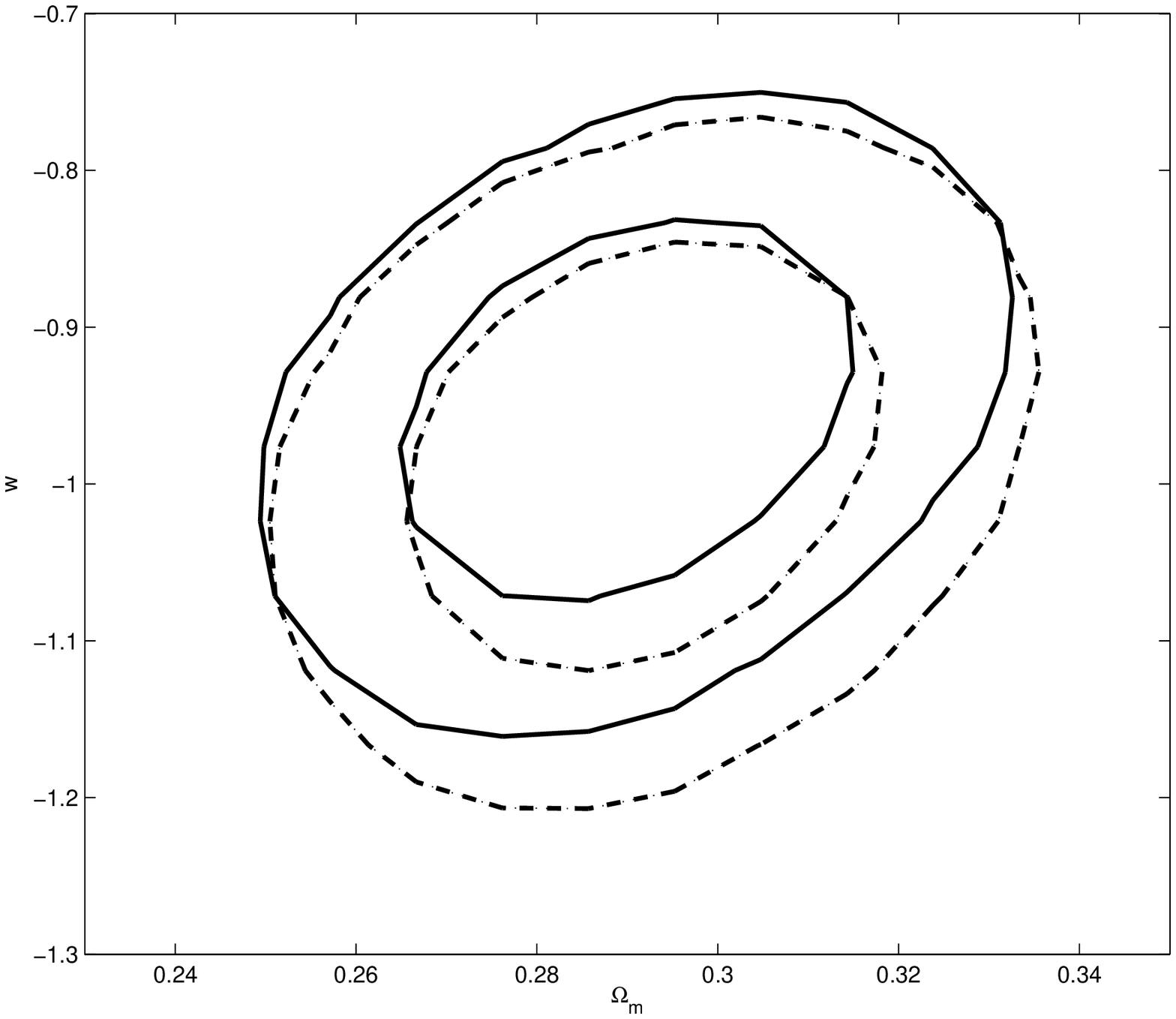}
\caption{1$\sigma$ and 2$\sigma$ $\Omega_{m}$-$w$ contours for the wCDM model. The solid lines 
correspond to the result of a combined analysis of the growth rate, SNIa, BAO and CMB data, 
while the dotted lines are obtained using only the SNIa, BAO and CMB data.}
\end{figure*} 

\begin{figure*}
\includegraphics[width=7in]{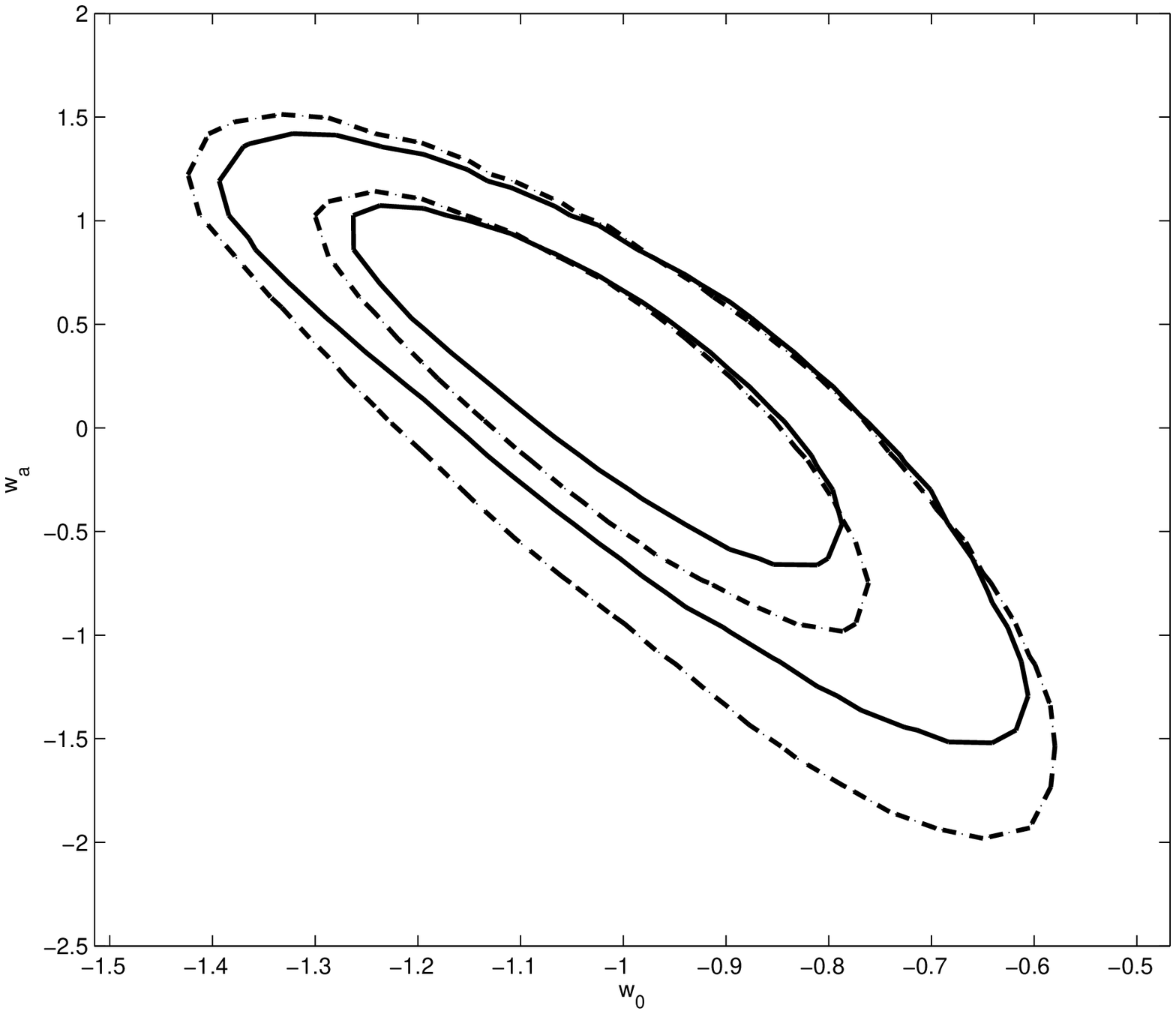}
\caption{1$\sigma$ and 2$\sigma$ $w_0$-$w_a$ contours for the CPL model. The solid lines 
correspond to the result of a combined analysis of the growth rate, SNIa, BAO and CMB data, 
while the dotted lines are obtained using only the SNIa, BAO and CMB data.}
\end{figure*}

\bibliographystyle{astroads}

\bibliography{growthratebib}







\end{document}